# Correlative approaches in single-molecule biophysics: a review of the progress in methods and applications

Mark C Leake[1,2]

[1] Department of Physics, University of York, U.K.

[2] Department of Biology, University of York, U.K.

## Abstract

Here, we discuss a collection of cutting-edge techniques and applications in use today by some of the leading experts in the field of correlative approaches in single-molecule biophysics.  A key difference in emphasis, compared with traditional single-molecule biophysics approaches detailed previously, is on the emphasis of the development and use of complex methods which explicitly combine multiple approaches to increase biological insights at the single-molecule level. These so-called correlative single-molecule biophysics methods rely on multiple, orthogonal tools and analysis, as opposed to any one single driving technique. Importantly, they span both *in vivo* and *in vitro* biological systems as well as the interfaces between theory and experiment in often highly integrated ways, very different to earlier traditional non-integrative approaches. The first applications of correlative single-molecule methods involved adaption of a range of different experimental technologies to the same biological sample whose measurements were synchronised. However, now we find a greater flora of integrated methods emerging that include approaches applied to different samples at different times and yet still permit useful molecular-scale correlations to be performed. The resultant findings often enable far greater precision of length and time scales of measurements, and a more understanding of the interplay between different processes in the same cell. Many new correlative single-molecule biophysics techniques also include more complex, physiologically relevant approaches as well as increasing number that combine advanced computational methods and mathematical analysis with experimental tools. Here we review the motivation behind the development of correlative single-molecule microscopy methods, its history and recent progress in the field.

## The value of correlative single-molecule biophysics

Since the middle of the last century, when the first grainy images of single filamentous biopolymers began to appear on transmission electron microscopy images (1), biophysics has been instrumental in helping to develop both experimental and theoretical techniques that can enable detection and quantification at the detection sensitivity of single biomolecules, and their associated interactions in complex processes (2)(3).  These methods enable substantive insights into the underlying mechanistic biology of cellular processes, but also can inform us about unexpected emergent details of the physics of life, one molecule at a time (4).

A crucial feature of biomolecules is their intrinsic *meta-stability* – the fact that they can adopt a range of different states of free energy separated by just a few thermal energy units of $k_BT$, where $k_B$ is the Boltzmann constant and $T$ the absolute temperature of the system. This explicit instability allows molecules to adopt different conformations coupled to biochemical processes that release chemical potential energy to bias thermodynamic transitions between different microstates, such as the hydrolysis of ATP. This is a critical reason why single-molecule biophysics approaches have so much value when studying many biological systems compared to more traditional ensemble average methods. They offer an ability to render a *probability distribution* for a physical and/or chemical characteristic. This approach utilises an implicit assumption of *ergodicity of molecular states* – namely, that any given single molecule under study will eventually sample all parts of that parameter space given sufficient time. Therefore, instead of reporting on the ensemble average output from several thousands of molecules, one can sample either the same single molecule, or several different molecules of the same type, over a sufficiently extended time to build up a near-complete picture of the underlying probability distribution relating to whichever molecular parameter is being measured. In essence, this renders information of not only the proportions of molecules in different microstates, but also what the related probabilities are for transitioning between these microstates. These quantitative details are core to understanding reaction kinetics and molecular dynamics of biological systems, the driving time-dependent processes of all cells.

That said, there are some rare exceptions in the life sciences for which ensemble level information can still render molecular-precise probability distributions. These are exemplified by instances of *population synchronicity*. For example, striated muscle tissue contains molecular motors of myosin molecules that undergo so-called power strokes that are not only synchronized, due to a signal influx from a rapid pulse of $Ca^{2+}$ ions, but also have population-level spatial periodicity across the extent of a muscle fiber due to a highly regular repeating functional unit of a single myofibril within the muscle tissue called the sarcomere. This means that ensemble level observations can be converted into molecular-scale information relatively trivially since there is little dephasing in signal either in time or space. It is no coincidence that both muscle-themed physiology and structural biology research were at the forefront of the development of modern biophysics and an ultimate distillation of a subset of techniques into the discipline of single-molecule biophysics.

The biggest limitation to single-molecule biophysics methods, whether experimental or theoretical, is ultimately sensitivity due to thermal physics. Since the heights of free energy activation barriers between different meta-stable microstates are comparable to the energy scale for thermal fluctuations in the surrounding water solvent molecules, this sets a challenge of detection confidence for any single-molecule measurement. Put another way, the signal-to-noise ratio for single-molecule detection using any single biophysics approach may often be only marginally above unity. Obvious ways to mitigate against this issue involve developing approaches that enhance the signal relative to the noise. The most common approach uses the fact that most sources of noise relevant to single-molecule detection are uncorrelated with time, whereas the signal is not. Exceptions to this include quantization noise, such as the noise created due to conversion from analog to digital signals, but in general other sources of uncorrelated detection noise dominate in single-molecule approaches. Therefore, if you extend sampling windows over longer times to generate a mean average output then you will reduce the noise but not the signal. This is simply *signal averaging* of course. A similar trick can be played with space, if a signal from a region of a sample is spatially extended, whereas the noise across that region will remain uncorrelated relative to its spatial location. In other words, one can generate a mean average signal from a greater region of space to improve the signal-to-noise ratio.

The principal drawback with these approaches is that what you have done is to sacrifice resolution in time or space, to gain confidence that what you have measured is indeed a true molecular signal as opposed to noise. This sacrifice is often the ultimate limit to our biological knowledge in many cellular processes. Key events in biological pathways occur either too rapidly, or over too small a distance, or both, such that their details remain shrouded in mystery due to the technical limitations of detection sensitivity. However, this is where correlative single-molecule biophysics approaches have a valuable role to play.

### Early applications of correlative single-molecule biophysics

A key benefit of correlating orthogonal signals from single molecules that are synchronised is that false-positive detection probabilities from each signal source are in effect multiplied resulting in an overall increase in the confidence of detection, namely that the associated measurement is indeed due to a single molecule under study and not to noise. For example, if the standard deviation noise from one detection device is σ and we set a signal detection threshold in subsequent analysis to be some small multiple of this to ensure we can detect a reasonable proportion of single-molecule events, say 2σ, then this might on initial inspection appear to be a high enough threshold to minimise noise detection. In fact, this still means that ~5%, or roughly 1 in 20 detected events, will be noise – a substantial proportion of the total data. If it is possible to have two similar detectors in operation, however, that are synchronised, and we insist that an event will only be flagged as "true" if detected at the same time on both detectors, then the total false positive probability will be 1/20 multiplied by 1/20. In other words, roughly 1 in 400 events marked as being putatively "single molecule" will be noise-related false positives. This strategy was used in a seminal paper by Smith and co-workers (5), in which they extended the capability of an earlier experimental *in vitro* assay in which two optically trapped microspheres bound either side of a single stiff F-actin filament were used to detect power stokes of the molecular motor myosin II (6). Unlike the earlier study, the positions of both microspheres were detected simultaneously. When the actin filament was lowered onto a third surface-immobilized microsphere covered in the active head subunit of myosin II molecules then the power stroke motions of this head on the actin track could be detected with significantly higher confidence. This led to improved estimates for the piconewton (pN) and nanometer (nm) scale forces and displacements respectively involved in the molecular conformational changes involved.

Arguably, the earliest application of correlating different biophysical measurements on the same single-molecule was the work of Perkins, Smith and Chu published in 1993, which involved correlating force translocation, in the form of fluid flow to extend a single DNA molecule tethered between a coverslip and a microsphere on the other end of the molecule, correlated with fluorescence microscopy of the labelled DNA, to enable observations of tube-like motions of the polymer chain (7). Other early applications of correlating force transduction methods with fluorescence microscopy on the same single-molecule at the same time have involved optical trapping and fluorescence microscopy, first demonstrated using dual-beam trapping with single-molecule total internal reflection fluorescence (smTIRF) to directly visualize the turnover of nucleotides during the molecular machine kinesin translocating on tubulin tracks (8). Similarly, in another study, magnetic tweezers were combined with smTIRF measurements, using single-molecule Förster resonance energy transfer (smFRET) to detect transitions in single-stranded DNA (9). Atomic force microscopy (AFM) force spectroscopy was later combined with TIRF to monitor conformational unfolding of single protein domains (10). AFM force spectroscopy was also correlated in another study with multiple analytical analysis tools, such as Monte Carlo simulations (11). Force transduction methods have also been combined at a far larger length scale relevant to whole cells, to mechanically perturb single cells whilst using optical super-resolution techniques to monitor the localization of key biomolecules (12).

Many, but not all, recent correlative biophysical methods have involved microscopy in some form (for an interesting overview see Ando et al 2018 (13), some of which also have relevance to single-molecule detection). Many of these do not necessarily rely on synchronisation of different techniques. For example, correlative light and electron microscopy (CLEM) (14) enables useful biological insights by utilising detection of specific fluorescently-labelled biomolecule components in light microscopy inside cells, and then viewing the same fixed cells using transmission electron microscopy to study high-resolution sub-cellular ultrastructure in the vicinity of these specific biomolecules. Cryo-fixation has proven invaluable in sample stabilization for CLEM, a technique that has assisted the development of super-resolution capability for the optical microscopy component (15). Methods using controllable freezing fixation that minimise artefacts associated with chemical fixation have also been applied to small X-ray scattering (SAXS) correlated with fluorescence microscopy that has permitted important insights that cross multiple length scales from the mesoscale of hundreds of nm ideal for SAXS down to a few tens of nm relevant to super-resolution fluorescence microscopy, such as in the new technique CryoSIM (16).

Electrical control and measurements have also been valuable to correlate with other biophysical parameters. For example, a rapidly rotating electric field can be used to rotate a dielectric particle attached to a biological rotary motor, whose speed of rotation can be monitored using a light microscopy method that utilises a weak optical trap in a laser interference position detection tool called back focal plane detection (17). Live cell measurements of transmembrane voltage using fluorescence microscopy reporter dyes can also be used to correlate the so-called protonmotive force (pmf) across bacterial cell membranes with the speed of rotation of the flagellar motor measured using back focal plane detection (18). Combining TIRF imaging with single-channel ion conductance measurements has similarly enabled important insights into the assembly mechanism of ion channels in lipid membranes (19).

The development of robust analytical frameworks for single-molecule biophysics tools has helped the roll-out of several new correlative approaches in particular involving fluorescence microscopy (20–22).  These have enabled methods to correlate the number of protein subunits in a diffusing molecular complex with its rate of diffusion measured from single-particle tracking of fluorescent protein reporters, for example on a key topisomerase enzyme involved in resolving torsional stress in DNA replication in bacteria (23), integrating the effects of multiple proteins used in bacterial cell division (24), and determining correlative effects between the size and shape of macromolecular protein structures such as carboxysomes found in photosynthesising bacteria and their local $CO_2$ microenvironment (25).

### A next generation of correlative single-molecule biophysics tools

What is on the horizon for emerging correlative single-molecule biophysics techniques? Although there is still growth in developing multiple single-molecule biophysics technologies into the same instrument to enable synchronised measurements on the same sample, perhaps a more emergent trend recently has been towards integrating tools and/or computational analysis that do not necessarily involve explicit synchronisation of different signal sources, and in fact may even be from different samples. These developments in many cases parallel increasing relevance towards physiological environments. For example, developments from Chien-Jung Lo and co-workers now extend the fluorescence-based voltage measurements on single live *Escherichia coli* cells suggesting that the bacterial flagellar motor can be used as a single cell voltmeter and mechano-detector; in essence it has potential applications as a multimodal biosensor (26). Julie Biteen's team have developed a new analytical tool that uses non-parametric Bayesian statistics on single-particle trajectory datasets obtained from live cell light microscopy that enables correlation of multiple mobility states, the average diffusion coefficient of single molecules in that state, the fraction of single molecules in that state, localization noise, and the probability of transitioning between states (27). Izzy Jayasinghe and colleagues have made valuable progress is development of expansion microscopy to enable the integration of TIRF imaging of $Ca^{2+}$ signals with DNA-PAINT imaging of nanoscale receptors on living primary cells that permits feature extraction and image alignment between correlative datasets for detecting ensembles of $Ca^{2+}$ channels that ere locally activated (28).

The group of Mike Heilemann has made advances in correlating DNA nanotechnology with multiplexed super-resolution imaging to enable highly precise quantification of membrane protein expression levels, clustering and distances in live cells expressing fibroblast growth factor receptors (29). Sviatlana Shashkova and Mikael Andersson et al have utilised multispectral millisecond single-molecule Slimfield imaging (30) to track the aquaglyceroporin complexes in live budding yeast cells in order to correlate their molecular stoichiometry and turnover kinetics with changes in a range of extracellular conditions (31). Jack Shepherd and Sarah Lecinski and co-workers have used rapid single-molecule optical microscopy on live yeast cells in combination with FRET-based sensors of local molecular crowding to enable the quantification of diffusing molecular trajectories on a single-cell level for correlating subcellular processes with their physicochemical environments under cell stress (32). Adam Wollman and team are developing single-molecule fluorescence tracking analysis that can extract different molecular signature patterns depending on the cellular nutrient environment, that serves as a barcode of the gene regulatory state of the cells which can be correlated with cell growth characteristics (33).

Alice Pyne's team have been developing a comprehensive open source software that enables morphological analysis of AFM imaging of single DNA molecules, permitting automated correlations of different morphological parameters to act as molecular topology signatures (34). Steven Quinn and researchers have correlated ratiometric imaging of a reporter dye to determine $Ca^{2+}$ concentrations inside model lipid vesicles, with stepwise single-molecule photobleaching of membrane-integrated molecules in aggregated amyloid protein implicated in dementia, to determine the dependence between their molecular stoichiometry and pore formation in the vesicles (35).

Maria Dienerowitz and co-workers have utilised an Anti-Brownian ELectrokinetic Trap (ABEL trap) to combine smFRET with molecular confinement to enable very extended observation times of up to several seconds for a key DNA binding protein involved in DNA replication, to explore its multiple molecular conformation states (36). The team of Johannes Holbein have developed phasor-based single-molecule localization microscopy to engineer the point spread function (PSF) shape of single fluorescent reporter dye molecules, integrating hardware modules for deformable mirror control with software for analysing the PSF properties (37). Alex Holehouse and researchers have integrated single-molecule spectroscopy and simulations for the study intrinsically disordered proteins (38). And the team of Jorge Bernardino de la Serna have developed a multi-modal super-resolution fluorescence microscopy system that can enable multi-dimensional and spatiotemporal correlative imaging at the plasma membrane of live cells to determine the how lipids adapt across multiple length scales from nano to micro (39).

The next generation of correlative single-molecule tools is expanding rapidly. The key reason for this is not so much the development of new technologies that are then applied to different biological systems, but more the need to address unresolved biological questions motivating the development of a range of exciting new technologies for the future, both experimental and theoretical. So, the momentum for new progress is being driven not so much by technologists and developers looking for new applications, but rather the challenging nature of open biological questions demanding new technological innovation. A fantastically efficient way to generate new technologies is to combine separate existing approaches into a correlative framework that enables genuinely new functionality.

# References

1. M. C. Leake, *Single-molecule cellular biophysics*, Cambridge University Press (2010) https:/doi.org/10.1017/CBO9780511794421.

2. M. C. Leake, *Biophysics : tools and techniques,* CRC Press (2016) https://doi.org/10.1201/9781315381589.

3. H. Miller, Z. Zhou, J. Shepherd, A. J. M. Wollman, M. C. Leake, Single-molecule techniques in biophysics: a review of the progress in methods and applications. *Reports Prog. Phys.* **81**, 024601 (2017).

4. M. C. Leake, The physics of life: one molecule at a time. *Philos. Trans. R. Soc. Lond. B. Biol. Sci.* **368**, 20120248 (2013).

5. D. A. Smith, W. Steffen, R. M. Simmons, J. Sleep, Hidden-Markov methods for the analysis of single-molecule actomyosin displacement data: The variance-Hidden-Markov method. *Biophys. J.* **81**, 2795–2816 (2001).

6. J. T. Finer, R. M. Simmons, J. A. Spudich, Single myosin molecule mechanics: Piconewton forces and nanometre steps. *Nature* **368**, 113–119 (1994).

7. T. T. Perkins, D. E. Smith, S. Chu, Direct observation of tube-like motion of a single polymer chain. *Science* **264**, 819–822 (1994).

8. T. Funatsu, *et al.*, Imaging and nano-manipulation of single biomolecules. *Biophys. Chem.* **68**, 63–72 (1997).

9. H. Shroff, *et al.*, Biocompatible force sensor with optical readout and dimensions of 6 nm3. *Nano Lett.* **5**, 1509–14 (2005).

10. A. Sarkar, R. B. Robertson, J. M. Fernandez, Simultaneous atomic force microscope and fluorescence measurements of protein unfolding using a calibrated evanescent wave. *Proc. Natl. Acad. Sci. U. S. A.* **101**, 12882–6 (2004).

11. W. A. Linke, M. C. Leake, Multiple sources of passive stress relaxation in muscle fibres. *Phys. Med. Biol.* **49**, 3613–27 (2004).

12. H. Colin-York, *et al.*, Super-Resolved Traction Force Microscopy (STFM). *Nano Lett.* **16**, 2633–2638 (2016).

13. T. Ando, *et al.*, The 2018 correlative microscopy techniques roadmap. *J. Phys. D. Appl. Phys.* **51**, aad055 (2018).

14. T. J. Deerinck, *et al.*, Fluorescence photooxidation with eosin: A method for high resolution immunolocalization and in situ hybridization detection for light and electron microscopy. *J. Cell Biol.* **126**, 901–910 (1994).

15. G. Wolff, C. Hagen, K. Grünewald, R. Kaufmann, Towards correlative super-resolution fluorescence and electron cryo-microscopy. *Biol. Cell* **108**, 245–258 (2016).

16. M. A. Phillips, *et al.*, CryoSIM: super-resolution 3D structured illumination cryogenic fluorescence microscopy for correlated ultrastructural imaging. *Optica* **7**, 802 (2020).

17. A. D. Rowe, M. C. Leake, H. Morgan, R. M. Berry, Rapid rotation of micron and submicron dielectric particles measured using optical tweezers. *J. Mod. Opt.* **50**, 1539–1554 (2003).

18. C.-J. Lo, M. C. Leake, T. Pilizota, R. M. Berry, Nonequivalence of membrane voltage and ion-gradient as driving forces for the bacterial flagellar motor at low load. *Biophys. J.* **93**, 294–302 (2007).

19. A. J. Heron, J. R. Thompson, B. Cronin, H. Bayley, M. I. Wallace, Simultaneous measurement of ionic current and fluorescence from single protein pores. *J. Am. Chem. Soc.* **131**, 1652–1653 (2009).

20. M. C. Leake, Analytical tools for single-molecule fluorescence imaging in cellulo. *Phys. Chem. Chem. Phys.* **16**, 12635–47 (2014).

21. T. Lenn, M. C. Leake, Experimental approaches for addressing fundamental biological questions in living, functioning cells with single molecule precision. *Open Biol.* **2**, 120090 (2012).

22. S.-W. S.-W. Chiu, M. C. Leake, Functioning nanomachines seen in real-time in living bacteria using single-molecule and super-resolution fluorescence imaging. *Int. J. Mol. Sci.* **12**, 2518–42 (2011).

23. M. Stracy, *et al.*, Single-molecule imaging of DNA gyrase activity in living Escherichia coli. *Nucleic Acids Res.* **47**, 210–220 (2019).

24. V. A. Lund, *et al.*, Molecular coordination of Staphylococcus aureus cell division. *Elife* **7** (2018).

25. Y. Sun, A. J. M. Wollman, F. Huang, M. C. Leake, L. N. Liu, Single-organelle quantification reveals stoichiometric and structural variability of carboxysomes dependent on the environment. *Plant Cell* **31**, 1648–1664 (2019).

26. E. Krasnopeeva, U. E. Barboza-Perez, J. Rosko, T. Pilizota, C. J. Lo, Bacterial flagellar motor as a multimodal biosensor. *Methods* (2020) https:/doi.org/10.1016/j.ymeth.2020.06.012.

27. J. D. Karslake, *et al.*, SMAUG: Analyzing single-molecule tracks with nonparametric Bayesian statistics. *Methods* (2020) https:/doi.org/10.1016/j.ymeth.2020.03.008.

28. M. E. Hurley, *et al.*, A correlative super-resolution protocol to visualise structural underpinnings of fast second-messenger signalling in primary cell types. *Methods* (2020) https:/doi.org/10.1016/j.ymeth.2020.10.005.

29. M. S. Schröder, *et al.*, Imaging the fibroblast growth factor receptor network on the plasma membrane with DNA-assisted single-molecule super-resolution microscopy. *Methods* (2020) https:/doi.org/10.1016/j.ymeth.2020.05.004.

30. M. Plank, G. H. Wadhams, M. C. Leake, Millisecond timescale slimfield imaging and automated quantification of single fluorescent protein molecules for use in probing complex biological processes. *Integr. Biol.* **1**, 602–12 (2009).

31. S. Shashkova, M. Andersson, S. Hohmann, M. C. Leake, Correlating single-molecule characteristics of the yeast aquaglyceroporin Fps1 with environmental perturbations directly in living cells. *Methods* (2020) https:/doi.org/10.1016/j.ymeth.2020.05.003.

32. J. W. Shepherd, *et al.*, Molecular crowding in single eukaryotic cells: Using cell environment biosensing and single-molecule optical microscopy to probe dependence on extracellular ionic strength, local glucose conditions, and sensor copy number. *Methods* (2020) https:/doi.org/10.1016/j.ymeth.2020.10.015.

33. S. Shashkova, T. Nyström, M. C. Leake, A. J. M. Wollman, Correlative single-molecule fluorescence barcoding of gene regulation in Saccharomyces cerevisiae. *Methods* (2020) https:/doi.org/10.1016/j.ymeth.2020.10.009.

34. J. G. Beton, *et al.*, TopoStats – A program for automated tracing of biomolecules from AFM images. *Methods* (2021) https:/doi.org/10.1016/j.ymeth.2021.01.008.

35. L. Dresser, *et al.*, Amyloid-β oligomerization monitored by single-molecule stepwise photobleaching. *Methods* (2020) https:/doi.org/10.1016/j.ymeth.2020.06.007.

36. M. Dienerowitz, J. A. L. Howard, S. D. Quinn, F. Dienerowitz, M. C. Leake, Single-molecule FRET dynamics of molecular motors in an ABEL trap. *Methods* (2021) https:/doi.org/10.1016/j.ymeth.2021.01.012.

37. K. J. A. Martens, A. Jabermoradi, S. Yang, J. Hohlbein, Integrating engineered point spread functions into the phasor-based single-molecule localization microscopy framework. *Methods* (2020) https:/doi.org/10.1016/j.ymeth.2020.07.010.

38. J. J. Alston, A. Soranno, A. S. Holehouse, Integrating single-molecule spectroscopy and simulations for the study of intrinsically disordered proteins. *Methods* (2021) https:/doi.org/10.1016/j.ymeth.2021.03.018.

39. M. Bernabé-Rubio, M. Bosch-Fortea, M. A. Alonso, J. B. de la Serna, Multi-dimensional and spatiotemporal correlative imaging at the plasma membrane of live cells to determine the continuum nano-to-micro scale lipid adaptation and collective motion. *Methods* (2021) https:/doi.org/10.1016/j.ymeth.2021.06.007.